\documentclass[aps,superscriptaddress,prd,twocolumn,nofootinbib]{revtex4}

\usepackage{amsmath,bm}
\usepackage{graphicx}
\usepackage{epstopdf}
\usepackage{subfigure}
\usepackage{epsfig}
\usepackage{amsmath,amssymb,amsfonts}
\usepackage{multirow}
\usepackage{tabularx}
\usepackage{color}
\usepackage[utf8]{inputenc}
\usepackage{verbatim}
\usepackage{array}
\usepackage{marvosym}
\usepackage{float}
\def\bea#1\eea{\begin{align}#1\end{align}}

\newcommand{\bef}{\begin{figure}[!htb]\centering}
\newcommand{\eef}{\end{figure}}

\newcommand{\fig}[1]{Fig.~\ref{#1}}

\graphicspath{{}} 

\begin{document}
\title{Transverse momentum balance of dijets in Xe+Xe collisions at the LHC}
\thanks{This work was supported by the Guangdong Major Project of Basic and Applied Basic Research No. 2020B0301030008 and the National Natural Science Foundation of China with Project Nos.~11935007, 12035007, 12247127, and 12247132. China Postdoctoral Science Foundation supports S. Wang under project No. 2021M701279.}

\date{\today  \hspace{1ex}}

\begin{abstract}

We present a theoretical study of the medium modifications of the $p_{\rm T}$ balance ($x_{\rm J}$) of dijets in Xe+Xe collisions at $\sqrt{s_{\rm NN}}=5.44$ TeV. The initial production of dijets was carried out using the POWHEG+PYTHIA8 prescription, which matches the next-to-leading-order (NLO) QCD matrix elements with the parton shower (PS) effect. The SHELL model described the in-medium evolution of nucleus-nucleus collisions using a transport approach. The theoretical results of the dijet $x_{\rm J}$ in the Xe+Xe collisions exhibit more imbalanced distributions than those in the p+p collisions, consistent with recently reported ATLAS data. By utilizing the Interleaved Flavor Neutralisation, an infrared-and-collinear-safe jet flavor algorithm, to identify the flavor of the reconstructed jets, we classify dijets processes into three categories: gluon-gluon ($gg$), quark-gluon ($qg$), and quark-quark ($qq$), and investigated the respective medium modification patterns and fraction changes of the $gg$, $qg$, and $qq$ components of the dijet sample in Xe+Xe collisions. It is shown that the increased fraction of $qg$ component at a small $x_{\rm J}$ contributes to the imbalance of the dijet; in particular, the $q_1g_2$ (quark-jet-leading) dijets experience more significant asymmetric energy loss than the $g_1q_2$ (gluon-jet-leading) dijets traversing the QGP. By comparing the $\Delta \langle x_{\rm J}\rangle = \langle x_{\rm J} \rangle_{\rm pp} - \langle x_{\rm J} \rangle_{\rm AA}$ of inclusive, $c\bar{c}$ and $b\bar{b}$ dijets in Xe+Xe collisions, we observe $\Delta \langle x_{\rm J} \rangle_{\rm incl.}>\Delta \langle x_{\rm J} \rangle_{\rm c\bar{c}}>\Delta \langle x_{\rm J} \rangle_{\rm b\bar{b}}$.  Moreover, $\rho_{\rm Xe, Pb}$, the ratios of the nuclear modification factors of dijets in Xe+Xe to those in Pb+Pb, were calculated, which indicates that the yield suppression of dijets in Pb+Pb is more pronounced than that in Xe+Xe owing to the larger radius of the lead nucleus.
\end{abstract}

\keywords{Heavy-ion collisions, Quark-gluon plasma, Jet quenching, Transverse momentum balance}

\author{Yao Li}
\affiliation{Key Laboratory of Quark \& Lepton Physics (MOE) and Institute of Particle Physics,
 Central China Normal University, Wuhan 430079, China}

 \author{Shu-Wan Shen}
 \affiliation{Key Laboratory of Quark \& Lepton Physics (MOE) and Institute of Particle Physics,
  Central China Normal University, Wuhan 430079, China}

\author{Sa Wang}
\email{wangsa@mails.ccnu.edu.cn}
\affiliation{College of Science, China Three Gorges University, Yichang 443002, China}
\affiliation{Center for Astronomy and Space Sciences and Institute of Modern Physics, China Three Gorges University, Yichang 443002, China}
\affiliation{Guangdong-Hong Kong Joint Laboratory of Quantum Matter, Guangdong Provincial Key Laboratory of Nuclear Science, Southern Nuclear Science Computing Center, South China Normal University, Guangzhou 510006, China}

\author{Ben-Wei Zhang}
\email{bwzhang@mail.ccnu.edu.cn}
\affiliation{Key Laboratory of Quark \& Lepton Physics (MOE) and Institute of Particle Physics, Central China Normal University, Wuhan 430079, China}
\maketitle

\section{introduction}
\label{sec-int}

Ultra-relativistic heavy-ion collisions at the Large Hadron Collider (LHC) and Relativistic Heavy Ion Collider (RHIC) provide a unique arena for searching for a new form of nuclear matter, quark-gluon plasma (QGP), in which the degrees of freedom of the quarks and gluons in the protons and neutrons are released \cite{Wang:1992qdg, Gyulassy:2003mc, Mehtar-Tani:2013pia, Qin:2015srf, Wang:2001cs}. The strong interactions between the hard-scattered partons with the medium, referred to as the ``jet quenching'' phenomenon, open up new avenues to understand the properties of such a strongly-coupled quark matter \cite{JET:2013cls, Xie:2022ght, Zhang:2021xib, JETSCAPE:2020mzn} and test the fundamental theory of quantum chromodynamics (QCD) at the extremely hot and dense conditions \cite{Collins:1974ky, Cunqueiro:2021wls, Cao:2020wlm, Tang:2020ame, Shen:2020mgh, Gao:2020vbh, Liu:2022ewr}. In the past two decades, a series of tools have been extensively investigated to reveal this partonic strong interaction, such as the suppression factor $R_{\rm AA}$ of high-$p_{\rm T}$ hadron/jet \cite{ATLAS:2014ipv, Wang:1998bha, Bass:2008rv, Wang:2002ri, He:2018xjv, JETSCAPE:2021ehl}, the momentum asymmetry of dijets \cite{ATLAS:2010isq, CMS:2011iwn, Zhang:2007ja, Qin:2010mn, Kang:2018wrs, Senzel:2013dta, Xing:2012ii, Kang:2011bp, Gao:2016ldo, Renk:2012cb, Chen:2016cof, Dai:2018mhw}, correlations of the vector boson associated jets ($\gamma/Z^0$+jets) \cite{CMS:2012ytf, Neufeld:2010fj, Dai:2012am, Huang:2015mva, Yang:2022nei, Chen:2020tbl}, the global event geometry \cite{Chen:2020pfa, Kang:2023qxb} and the jet substructures~\cite{CMS:2013lhm, Vitev:2009rd, Vitev:2008rz, Zhang:2021sua, Chen:2022kic, Zhang:2018urd, Wang:2023eer, JETSCAPE:2023hqn, Wang:2022yrp, Kang:2023ycg}.

Because dijets are the dominant QCD processes in the hadron collisions in the experiment and are less influenced by the underlying background than inclusive jets, they exhibit a unique glamour in jet physics. The back-to-back configurations of the two leading jets in the transverse plane can significantly suppress the contribution of the underlying background of the jet reconstruction in both p+p and A+A collisions. In a vacuum, the parton shower effects and higher-order QCD processes may break the symmetry of the final-state dijets, leading to a deflection from the back-to-back azimuthal angle and an unequal transverse momentum between the leading and subleading jets. In A+A collisions, because the two jets usually experience asymmetric energy loss as traversing the QGP medium, the transverse momentum balance of dijets $x_{\rm J} \equiv p_{\rm T,2}/p_{\rm T,1}$ \cite{ATLAS:2017xfa}, defined as the ratio of the subleading to leading jet $p_{\rm T}$, can be further modified by the in-medium interactions and show a different sensitivity to the path-length dependence of jet quenching \cite{Qin:2010mn} and jet-by-jet fluctuations of jet-medium interactions \cite{Milhano:2015mng}. More imbalanced $A_{\rm J} \equiv (p_{\rm T,1} - p_{\rm T,2})/(p_{\rm T,1} + p_{\rm T,2})$ and $x_{\rm J}$ distributions of dijets have been observed in Pb+Pb collisions relative to p+p at $\sqrt{s_{\rm NN}}=2.76$ TeV and $\sqrt{s_{\rm NN}}=5.02$ TeV by the ATLAS \cite{ATLAS:2010isq,ATLAS:2017xfa,ATLAS:2022zbu} and CMS Collaborations \cite{CMS:2011iwn,CMS:2012ulu,CMS:2018dqf},
and in Au+Au collisions at $\sqrt{s_{\rm NN}}=200$ GeV by the STAR Collaborations~\cite{STAR:2016dfv},
which have been extensively investigated using theoretical calculations \cite{Kang:2018wrs, Senzel:2013dta, Gao:2016ldo, Renk:2012cb, Chen:2016cof, Dai:2018mhw}.

Recently, the ATLAS Collaboration measured the dijet $x_{\rm J}$ in Xe+Xe collisions at $\sqrt{s_{\rm NN}}=5.44$ TeV for the first time~\cite{ATLAS:2023xzy} however, timely theoretical studies are still lacking. Because the xenon nucleus has a smaller radius than lead, studying the jet production in different collision systems will deepen our understanding of the system size dependence of the jet quenching effect~\cite{ALICE:2018hza,ATLAS:2019dct, ATLAS:2022zbu, ATLAS:2017xfa, Xie:2019oxg, Li:2021xbd, Zhang:2022fau}. Furthermore, because dijet events consist of $gg$, $qg$ and $qq$ components, whereas the jet energy loss is closely related to the flavor of hard partons ($\Delta E_g/\Delta E_q\sim C_A/C_F = 9/4, C_A = 3, C_F = 4/3$) \cite{Gyulassy:2003mc, Wang:1998bha}, it is essential to determine their respective modification patterns and assess the roles they play in the overall medium modifications of dijet $x_{\rm J}$. Furthermore, massive heavy quarks are believed to lose less energy than light quarks owing to the ``dead-cone'' effect \cite{Dokshitzer:2001zm, Zhang:2003wk, Armesto:2003jh, Djordjevic:2003qk}, which leads to a mass hierarchy of partonic energy loss $\Delta E_q>\Delta E_c>\Delta E_b$ \cite{Xing:2019xae, Xing:2023ciw, Wang:2023eer, Zhang:2023oid}. It is of particular interest to explore the mass dependence of the medium modification on the dijet $x_{\rm J}$ by comparing light- and heavy-flavor (such as $c\bar{c}$ and $b\bar{b}$) dijets in high-energy nuclear collisions.

This paper presents the first theoretical study on medium modifications of the dijet $p_{\rm T}$–balance $x_{\rm J}$ in Xe+Xe collisions. The initial production of dijets was carried out using the POWHEG+PYTHIA8 prescription, which matches the next-to-leading-order (NLO) QCD matrix elements with the parton shower (PS) effect. The transport approach describes dijets' in-medium evolution, which considers both elastic and inelastic partonic interactions in the quark-gluon plasma (QGP). First, we present the theoretical results of the dijet $x_{\rm J}$ in Xe+Xe collisions at $\sqrt{s_{\rm NN}}=5.44$ TeV compared with recently reported ATLAS measurements. Specifically, we discuss medium modification's flavor and mass dependence on the dijet $x_{\rm J}$. We studied the respective medium-modification patterns and fraction changes of the $gg$, $qg$, $qq$, as well as the $q_1g_2$ and $g_1q_2$ components in the dijet samples for both p+p and Xe+Xe collisions. We also investigated the mass effect of the $x_{\rm J}$ modifications by comparing the $\Delta x_{\rm J}$ of the inclusive, $c\bar{c}$ and $b\bar{b}$ dijets in the Xe+Xe collisions. Finally, we present the calculated results of dijets nuclear modification factor for Xe+Xe at $\sqrt{s_{\rm NN}}=5.44$ TeV and Pb+Pb at $\sqrt{s_{\rm NN}}=5.02$ TeV compared with recent ATLAS data.

\section{Theoretical framework}
\label{framework}

In this study, we generated next-to-leading-order (NLO) matrix elements for QCD jet processes \cite{Alioli:2010xa} in the framework of POWHEG-BOX-V2 \cite{Nason:2004rx, Frixione:2007vw, Alioli:2010xd} and then simulated the parton shower (PS) with PYTHIA 8.309 \cite{Bierlich:2022pfr} to produce p+p events. CT18NLO parton distribution functions (PDF) \cite{Hou:2019qau} were chosen for computation. The jets were reconstructed using the anti-$k_{\rm T}$ clustering algorithm and the radius parameter $R$ = 0.4, as implemented in the FastJet package~\cite{Cacciari:2011ma}. Then, the two highest $p_{\rm T}$ jets out of the set of jets in an event are selected as the dijet candidates. The leading jet transverse momentum $p_{\rm T,1}$ and subleading jet transverse momentum $p_{\rm T,2}$ must be greater than 100 GeV/c and 32 GeV/c, respectively. The two jets are required to be nearly back-to-back in azimuth with $\Delta \phi \equiv |\phi_{1} - \phi_{2}| \geq 7\pi/8$ and to be in the rapidity region $|y| < 2.1$. If all these conditions are satisfied, the desired dijet candidate is accepted.

\bef
\includegraphics[width=0.9\linewidth]{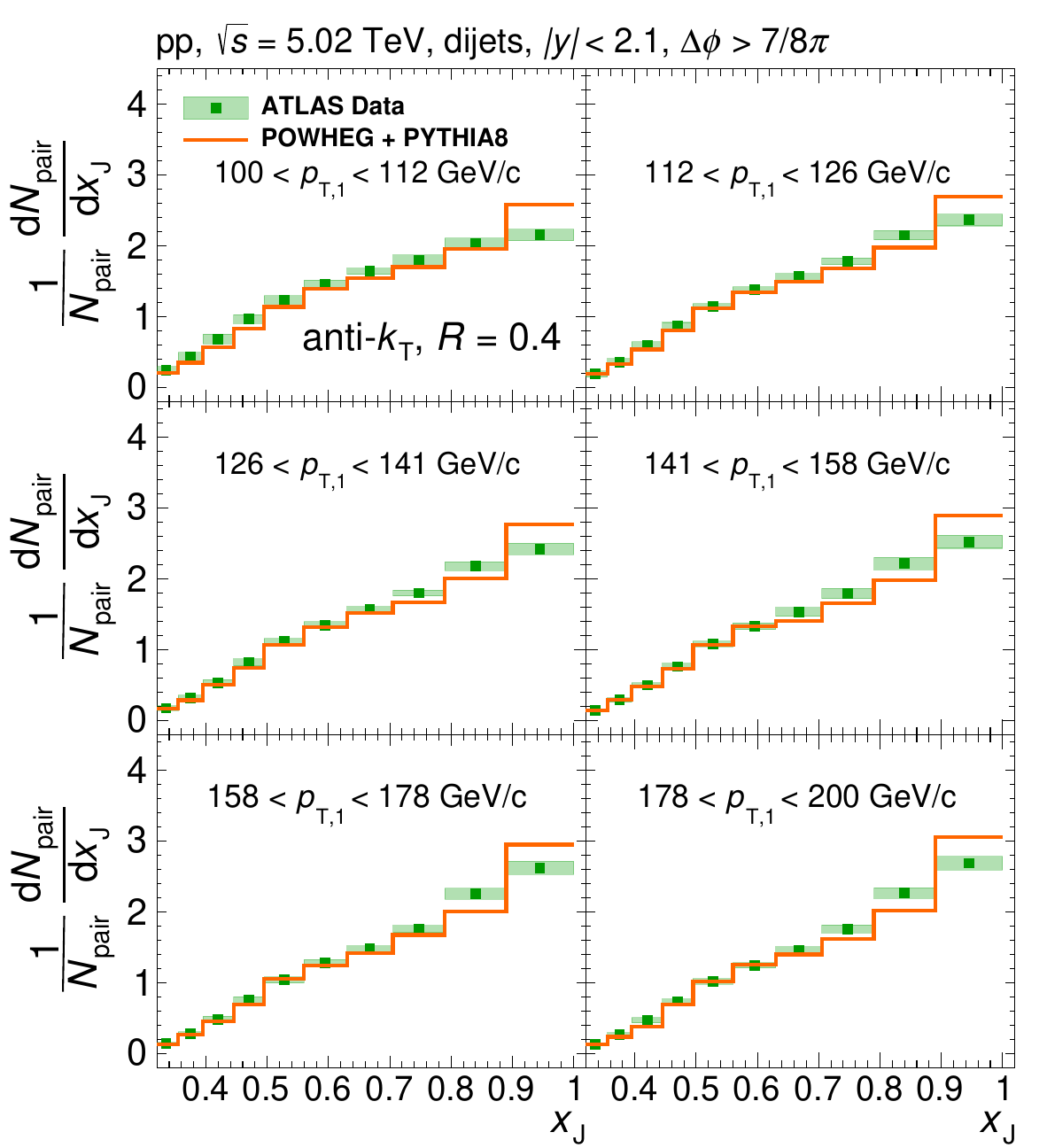}
\caption{ (Color online) Normalized $x_{\rm J}$ distributions of dijets in p+p collisions at $\sqrt{s}=5.02$ TeV for six $p_{\rm{T,1}}$ intervals: [100, 112], [112, 126], [126, 141], [141, 158], [158, 171] and [171, 200] GeV/c are compared with the ATLAS data~\cite{ATLAS:2022zbu}.}
\label{fig:xj_5020_baseline}
\eef

We calculated the normalized $x_{\rm J}$ distributions in $ p+p $ collisions and compared them with the ATLAS data~\cite{ATLAS:2022zbu} as shown in \fig{fig:xj_5020_baseline}. We observe that the $x_{\rm J}$ distributions calculated by POWHEG+PYTHIA8 provide decent descriptions at small value of $x_{\rm J}$ for all six different $p_{\rm{T,1}}$ intervals compared to the ATLAS data, except for a slight overestimation of $x_{\rm J}$ distributions compared to the ATLAS data near 1. At each $p_{\rm T}$ interval, the $x_{\rm J}$ distribution peaks near $x_{\rm J} \simeq 1$, where the leading and subleading jets are almost balanced. However, higher-order perturbative QCD corrections and splitting processes during the parton shower in a vacuum produce a considerable fraction of dijets with an imbalanced transverse momentum in the smaller $x_{\rm J}$ region.

The in-medium evolution of both light- and heavy-flavor diets was simulated using the SHELL model \cite{Dai:2018mhw, Wang:2020ukj}, which considers the elastic and inelastic partonic energy loss within the hot/dense QGP medium. The initial spatial distribution of hard scatterings was sampled using a Monte Carlo Glauber model~\cite{Miller:2007ri}. Since the propagation of massive partons in the QCD medium can be viewed as ``Brownian motion," the transport of heavy quarks can be well described by the modified Langevin equations,

\begin{eqnarray}
&&\Delta\vec{x}(t)=\frac{\vec{p}(t)}{E}\Delta t,\\
&&\Delta\vec{p}(t)=-\eta_D\vec{p}\Delta t+\vec{\xi}(t)\Delta t-\vec{p}_{\rm g}(t) .
\label{eq:lang2}
\end{eqnarray}

These two equations describe heavy quarks' position and momentum updates traversing the QGP medium. $\eta_D$ is the drag coefficient that controls the energy-dissipation strength of heavy quarks in the medium. The stochastic term $\xi(t)$ denotes random kicks as heavy quarks scattered by a thermal particle, which obeys a Gaussian distribution. The diffusion coefficient $\kappa$ can be related to $\eta_D$ using the fluctuation-dissipation theorem $\kappa=2\eta_D ET$. The first two terms on the right-hand side of Eqs. ~\ref{eq:lang2} represents the collisional energy loss of heavy quarks. In contrast, the last term, $-p_g$ is the momentum correction caused by medium-induced gluon radiation. In our framework, the higher-twist \cite{Guo:2000nz, Zhang:2003wk, Zhang:2003yn, Majumder:2009ge} formalism was employed to simulate the medium-induced gluon radiation of the jet partons, in which the gluon radiation spectra of an energetic parton in the QGP can be obtained as

\begin{eqnarray}
\frac{dN_g}{ dxdk^{2}_{\perp}dt}=\frac{2\alpha_{s}P(x)\hat{q}}{\pi k^{4}_{\perp}}\sin^2(\frac{t-t_i}{2\tau_f})(\frac{k^2_{\perp}}{k^2_{\perp}+x^2M^2})^4,
\label{eq:dndxk}
\end{eqnarray}
where $x$ and $k_{\perp}$ denote the energy fraction and transverse momentum of the radiated gluon, respectively. $P(x)$ is the QCD splitting function for the splitting processes $g\rightarrow g+g$ and $q(Q)\rightarrow q(Q)+g$ ~\cite{Wang:2009qb}, $\tau_f=2Ex(1-x)/(k^2_\perp+x^2M^2)$ is the formation time of the daughter gluon. $\hat{q}$ denotes the general jet transport parameter in QGP~\cite{Chen:2010te}. The last term in Eq.~\ref{eq:dndxk} represents the suppression factor resulting from the ``dead-cone'' effect of heavy quarks \cite{Dokshitzer:2001zm, ALICE:2019nuy}, which reduces the probability of gluon radiation within a small cone ($\theta\sim M_Q/E$). The collisional energy loss is generally dominant for low-energy heavy quarks due to the ``dead-cone'' effect. In contrast, radiative energy loss is usually expected to become significant at $p_{\rm T}^Q > 5m_Q$ \cite{Dong:2019byy}.

For massless partons, the collisional energy loss is estimated by pQCD calculations within the Hard-Thermal Loop approximation \cite{Neufeld:2010xi, Huang:2013vaa}, whereas the same higher-twist formalism estimates their radiative contribution as for massive partons. The hydrodynamic time-space evolution of the QGP medium is described using CLVisc programs~\cite{Pang:2016igs, Pang:2018zzo}, which provide the temperature and velocity of the expanding hot/dense nuclear matter. The SHELL model has been successfully applied in the study of heavy-flavor jets in high-energy nuclear collisions, which gives satisfactory descriptions on a series of experiment measurements, such as $p_{\rm T}$ imbalance \cite{Dai:2018mhw}, radial profiles \cite{Wang:2019xey, Wang:2020ukj, Wang:2020ffj, Wang:2020bqz} and fragmentation functions \cite{Li:2022tcr} of heavy-flavor jets, correlations of $Z^0+$ HF hadron/jet \cite{Wang:2020qwe, Wang:2021jgm}.

\bef
\subfigure[]{
\label{fig:xj_5440_baseline}
\includegraphics[width=0.9\linewidth]{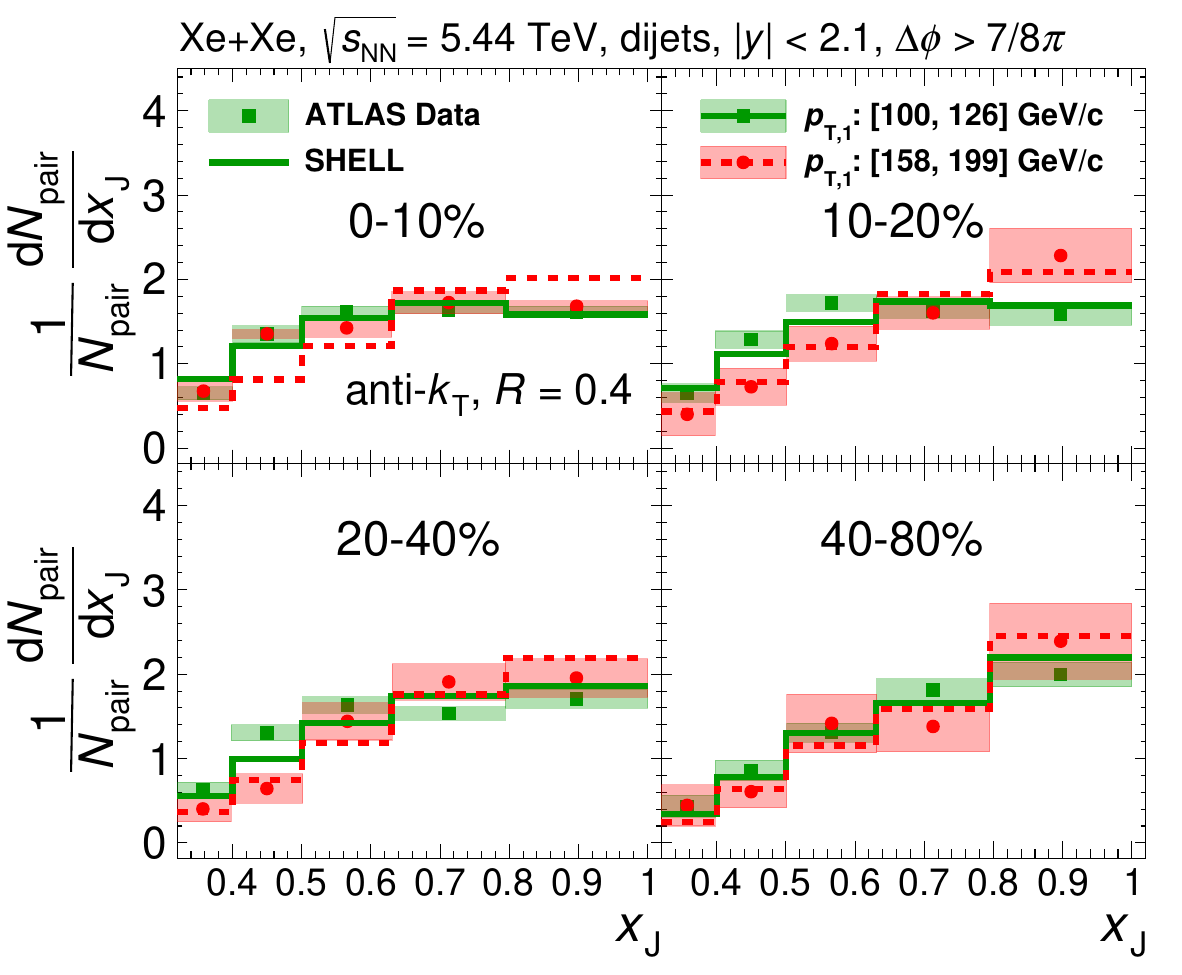}}

\subfigure[]{
\label{fig:xj_5440_RAA}
\includegraphics[width=0.9\linewidth]{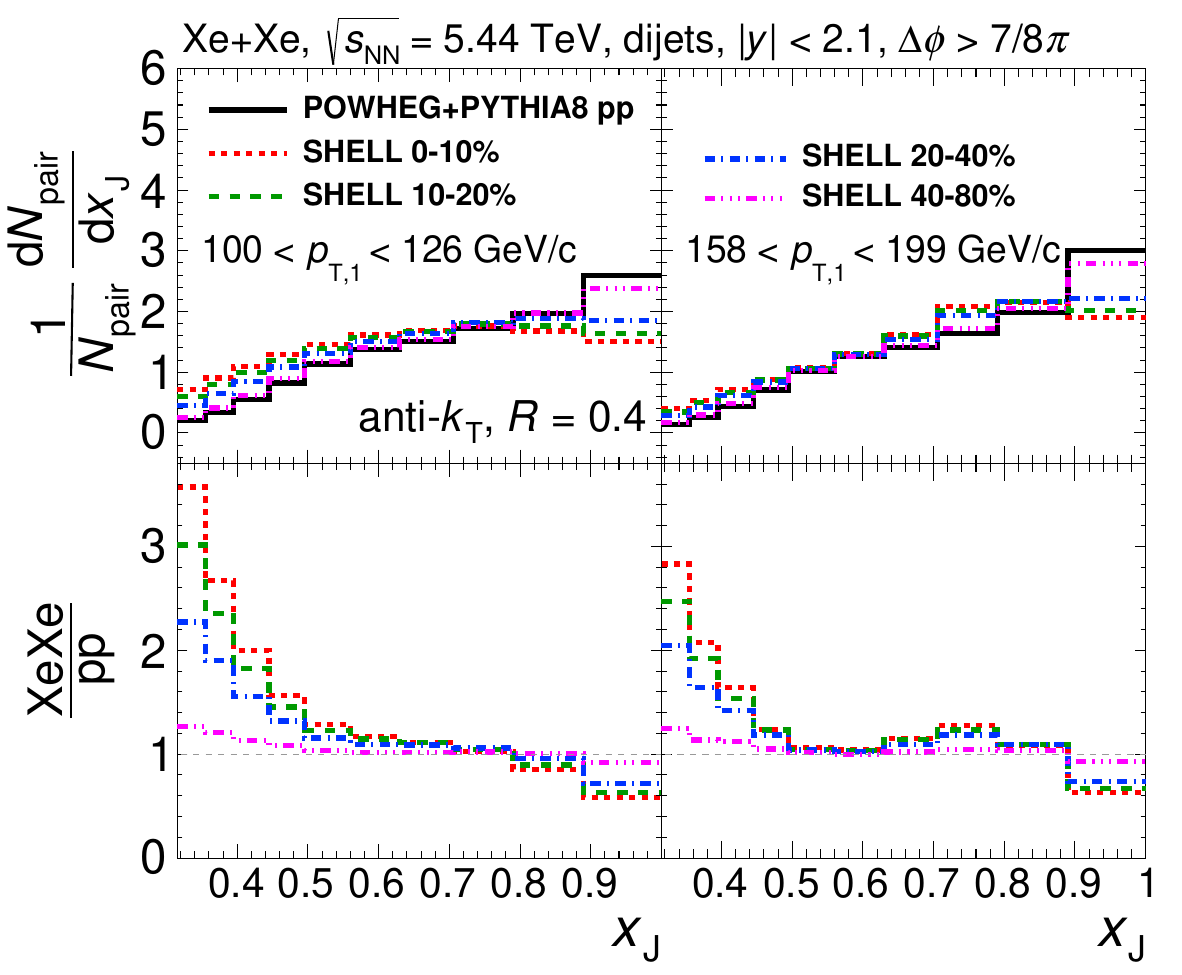}}
\caption{ (Color online) (a) Calculated normalized dijet $x_{\rm J}$ distributions in Xe+Xe collisions at $\sqrt{s_{\rm NN}}=5.44$ TeV are compared with the ATLAS data for four centrality bins (0\%-10\%, 10\%-20\%, 20\%-40\% and 40\%-80\%) and two $p_{\rm{T,1}}$ intervals (green: [100, 126] GeV/c, red: [158, 199] GeV/c). (b) Comparison of the normalized dijets $x_{\rm J}$ distributions between the Xe+Xe and p+p collisions at $\sqrt{s_{\rm NN}}=5.44$ TeV for two $p_{\rm{T,1}}$ intervals (left: [100, 126] GeV/c, right: [158, 199] GeV/c) and the ratios of XeXe/pp are shown in the bottom panels.}
\eef

\begin{table*}[htb]
    \centering
    \begin{tabular}{|c|c|c|c|c|c|c|}
    \hline
    \multirow{2}{*}{} & \multicolumn{3}{c|}{Pb+Pb}&\multicolumn{3}{c|}{Xe+Xe} \\
    \cline{2-7}

    & \quad $b_{\rm min}$ (fm) & \quad $b_{\rm max}$ (fm) & \quad $\langle b \rangle$ (fm)
    & \quad $b_{\rm min}$ (fm) & \quad $b_{\rm max}$ (fm) & \quad $\langle b \rangle$ (fm) \\
    \hline

    0\%-10\% & 0 & 4.98 & 3.52 & 0 & 4.28 & 3.02 \\
    \hline

    10\%-20\% & 4.98 & 7.05 & 6.10 & 4.28 & 6.05 & 5.24 \\
    \hline

    20\%-40\% & 7.05 & 9.98 & 8.64 & 6.05 & 8.56 & 7.41 \\
    \hline

    40\%-80\% & 9.98 & 14.11 & 12.22 & 8.56 & 12.12 & 10.49 \\
    \hline
    \end{tabular}
    \caption{Relations between centrality bins and the values of impact parameter $b$ for
    $\sqrt{s_{\rm NN}} = 5.02$ TeV Pb+Pb 
    and $\sqrt{s_{\rm NN}} = 5.44$ TeV Xe+Xe 
    collisions calculated by Glauber model.}
    \label{tab:impact_b}
\end{table*}

\section{Numerical Results and Discussions}
In \fig{fig:xj_5440_baseline}, we firstly show the dijet $x_{\rm J}$ distributions in Xe+Xe collisions at $\sqrt{s_{\rm NN}}=5.44$ TeV calculated by the SHELL model as a comparison to the ATLAS data for four centrality bins (0\%-10\%, 10\%-20\%, 20\%-40\% and 40\%-80\%) and two $p_{\rm{T,1}}$ intervals ([100, 126] GeV/c and [158, 199] GeV/c).
The centrality bins corresponding to the values of $b$ are listed in Table~\ref{tab:impact_b}, which are calculated using a Monte Carlo Glauber model~\cite{Miller:2007ri}.
We use the average impact parameter $\langle b \rangle$ in our calculations for simplicity.
We found that the calculations using the SHELL model provide a satisfactory description of the recently reported ATLAS data for almost all four centrality bins and two $p_{\rm{T,1}}$ intervals, except for the data at $158 < p_{\rm{T,1}} < 199$ GeV/c in the central 0\%-10\% collisions, which shift the $x_{\rm J}$ distribution towards larger $x_{\rm J}$ values compared with the ATLAS data. Furthermore, our theoretical results show a more balanced $x_{\rm J}$ distribution for higher $p_{\rm T}$ dijets ($158 < p_{\rm{T,1}} < 199$ GeV/c) than for lower ones, which is consistent with the trend observed in the ATLAS measurements. To determine the collision centrality and jet $p_{\rm T}$ dependence of the medium modification on the dijet $x_{\rm J}$ distributions, we compared the dijet $x_{\rm J}$ distributions in the Xe+Xe collisions with their p+p baseline for two $p_{\rm T}$ intervals, as shown in the bottom panels of \fig{fig:xj_5440_RAA}. The $x_{\rm J}$ distributions were observed to shift towards smaller $x_{\rm J}$ values in the Xe+Xe collisions compared with their p+p baseline, especially for the most central collisions. The phenomenon in which the dijet transverse momentum becomes more imbalanced in A+A collisions owing to the asymmetric energy loss suffered by the leading and subleading jets as the dijet traverses the QGP \cite{Qin:2010mn, Milhano:2015mng} has been observed in previous measurements of Pb+Pb collisions at 2.76 TeV \cite{ATLAS:2017xfa} and 5.02 TeV \cite{ATLAS:2022zbu}. In central collisions, such an asymmetric energy loss between the leading and subleading jets can be more significant because of the larger medium size and higher temperature than in the peripheral case. In addition, we observe a slightly weaker $x_{\rm J}$ modification for higher $p_{\rm T}$ dijets than for lower ones, which is consistent with expectations.

\bef
\includegraphics[width=0.9\linewidth]{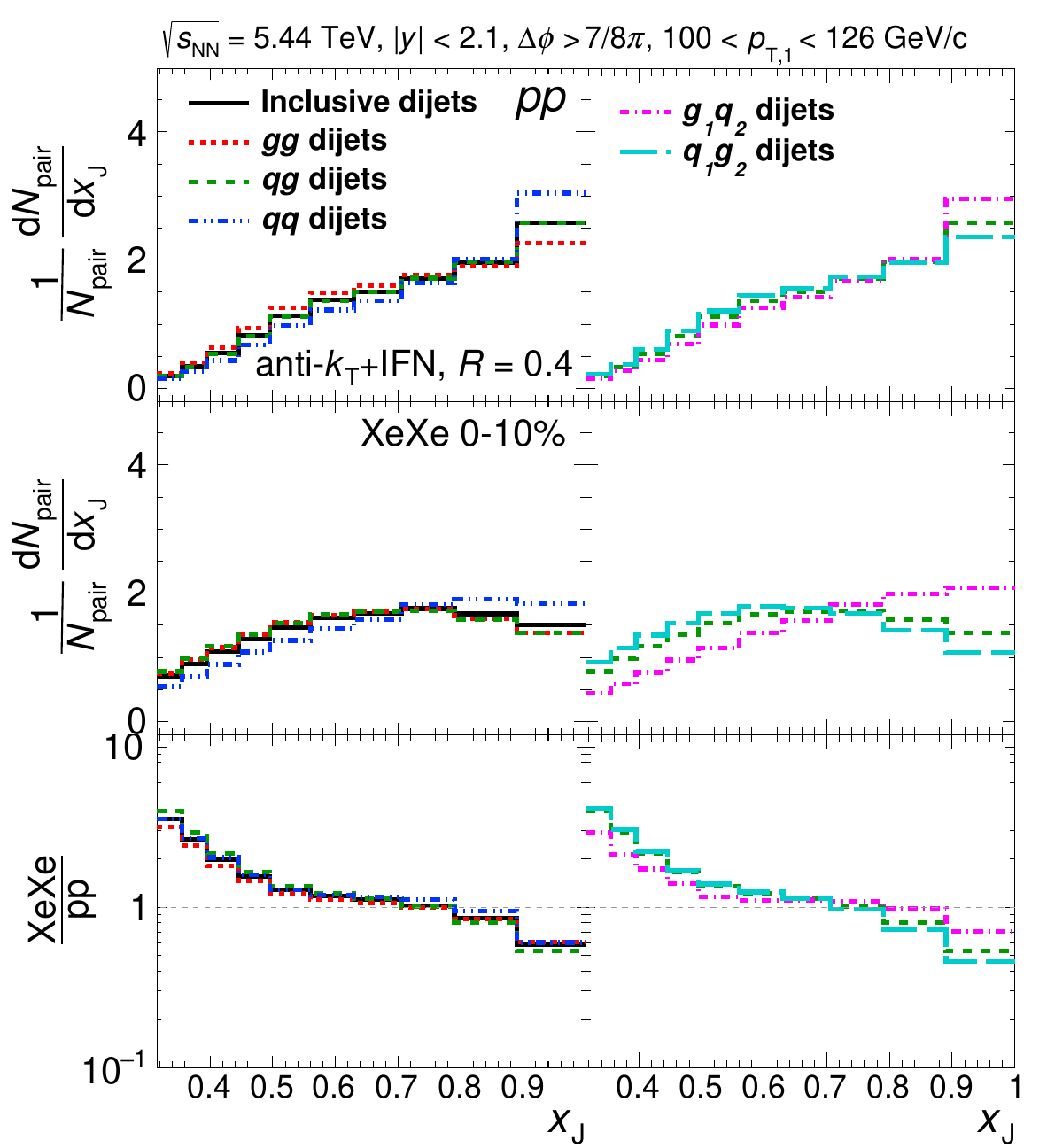}
\caption{ (Color online) Calculated normalized $x_{\rm J}$ distributions in p+p (top) and 0\%-10\% Xe+Xe (middle) collisions as well as their ratios (bottom) for inclusive, gluon-gluon, quark-gluon, and quark-quark dijets (left column); quark-jet-leading and gluon-jet-leading of quark-gluon dijets (right column). The $x_{\rm J}$ distributions of quark-gluon dijets (dashed green line) in the right column are identical to those in the left column. }
\label{fig:xj_5440_class}
\eef

The flavor dependence of the medium modification of dijet $p_{\rm T}$ balance is also an interesting topic in heavy ion collisions. Because of the different color factors,
the gluon-initiated jets are expected to lose more energy than the quark-initiated one as traversing the hot and dense nuclear matter, and the ``dead-cone'' effect also leads to smaller energy loss of heavy quarks than that of the light one \cite{Gyulassy:2003mc, Wang:1998bha,Dokshitzer:2001zm, Zhang:2003wk, Armesto:2003jh, Djordjevic:2003qk}. An optimal step toward this goal is determining the flavor of the selected jets in nucleus-nucleus collisions using an infrared and collinear (IRC) safe jet algorithm. In recent years, four IRC-safe jet-algorithm-based approaches have been developed to identify the flavor of the underlying hard partons for the final-state reconstructed jets, called ``flavor-$k_{\rm T}$'' \cite{Banfi:2006hf}, ``flavor anti-$k_{\rm T}$'' \cite{Czakon:2022wam}, ``flavor-dressing'' \cite{Gauld:2022lem} and ``Interleaved Flavor Neutralisation (IFN)'' \cite{Caola:2023wpj}. In this study, we employ the IFN algorithm to identify the flavor of the reconstructed jets at the parton level for gluon jets, light-quark jets, and heavy-flavor jets in both p+p and nucleus-nucleus collisions. Because the flavor information of the jets is accessible at each stage of the clustering sequence, the IFN algorithm provides consistent flavor identification for both the full jets and their substructures. It should be noted that identifying gluon and quark jets in experiments remains challenging, although such identification can be achieved through Monte Carlo simulations based on the complete flavor information of the jet particles. Note that the IFN algorithm is applicable to all jets at the parton level but is currently only accessible for heavy-flavor jets at the hadron level. As the first theoretical exploration, in the present work, we identified the jet flavor at the parton level; hence, flavor identification of both inclusive dijets and heavy-flavor dijets could be efficiently implemented.

Using the IFN algorithm, inclusive dijet events can be classified into three categories: gluon-gluon ($gg$), quark-gluon ($qg$), and quark-quark ($qq$). In particular, by distinguishing the flavor of the leading jet, $qg$ dijets can be further divided into $q_1g_2$ and $g_1q_2$, denoting the quark-jet-leading and gluon-jet-leading quark-gluon dijets, respectively. In the left column of \fig{fig:xj_5440_class}, we show the normalized $x_{\rm J}$ distributions of $gg$, $qg$, $qq$, and inclusive dijets in both p+p and 0\%-10\% Xe+Xe collisions at $\sqrt{s_{\rm NN}}=5.44$ TeV, and the ratio of XeXe/pp is plotted in the bottom panel. We find that $gg$ dijets have a more imbalanced initial distribution than $qq$. Quarks and gluons experience different parton shower processes in a vacuum because of their different color factors and splitting functions \cite{Bierlich:2022pfr, Sjostrand:2006za}. In the bottom panel, the $qg$ dijets exhibit a slightly stronger suppression near $x_{\rm J} \sim 1$ and enhancement near $x_{\rm J} \sim$ 0 than the others. The different color charges carried by the leading and subleading jets of $qg$ dijets lead to enhanced asymmetric energy loss in Xe+Xe collisions relative to the $qq$ and $gg$ jets. However, $qg$ dijets contain two types of subsets, $q_1g_2$ and $g_1q_2$, which may exhibit different modification patterns. In the right column of \fig{fig:xj_5440_class}, we observe a stronger enhancement at small $x_{\rm J}$ and stronger suppression at $x_{\rm J}\sim$1 on the $x_{\rm J}$ distribution of $q_{1}g_{2}$ dijets than that of $g_{1}q_{2}$. Compared to the $g_{1}q_{2}$ dijets, the leading jet of $q_{1}g_{2}$ dijets lost less energy, whereas the subleading jet lost more energy. In other words, the flavor configuration of $q_{1}g_{2}$ dijets results in a more significant asymmetric energy loss than $g_{1}q_{2}$ when traversing the QGP medium.

\bef
\includegraphics[width=0.9\linewidth]{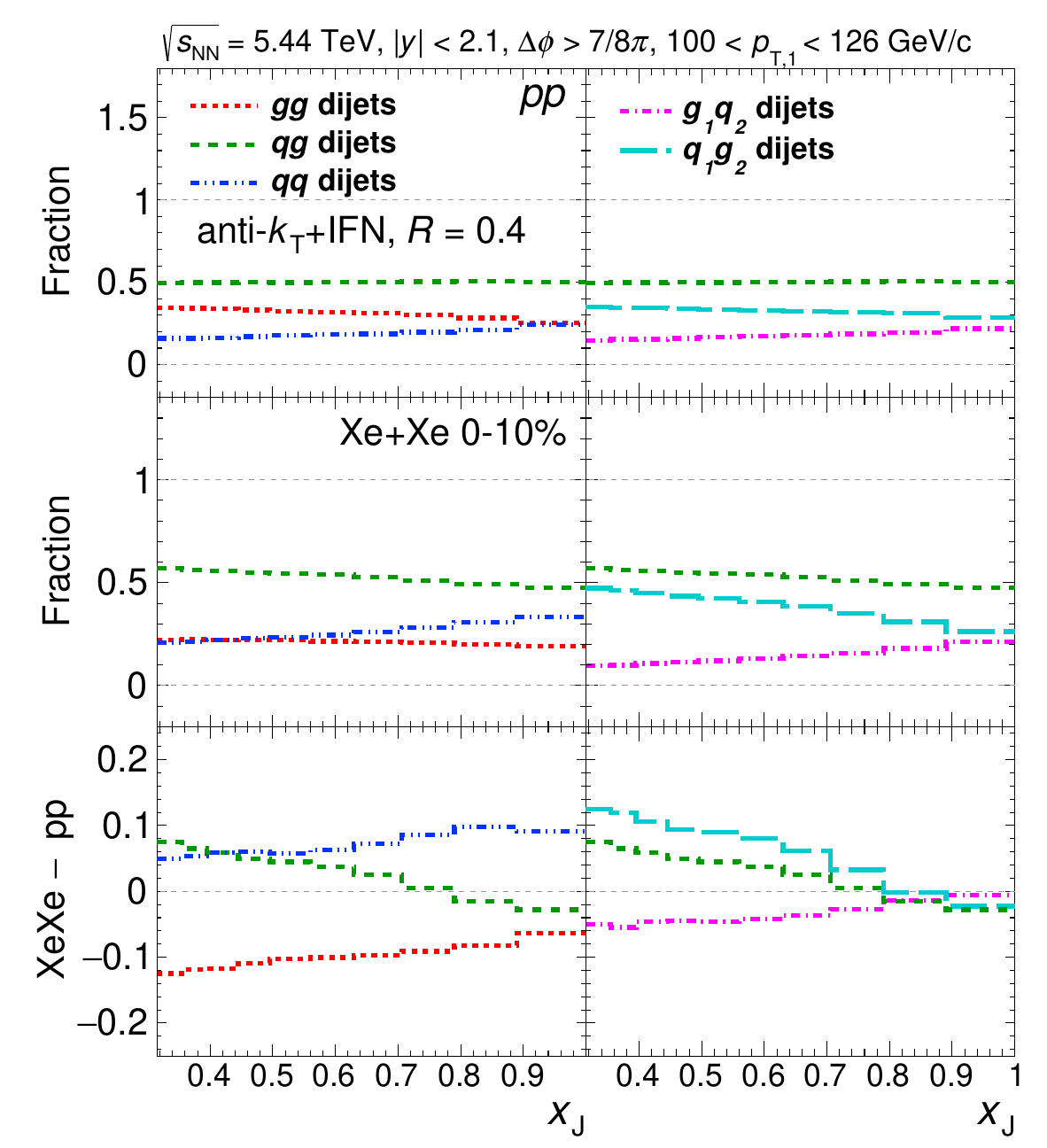}
\caption{ (Color online) Calculated fraction of subset dijets in inclusive dijets as the function of $x_{\rm J}$ in p+p (top) and 0\%-10\% Xe+Xe (middle) collisions as well as their differences (bottom) for gluon-gluon, quark-gluon, and quark-quark dijets (left column); quark-jet-leading and gluon-jet-leading quark-gluon dijets (right column). The fraction distributions of the quark-gluon dijets in the right column (dashed green line) are the same as those in the left column. }
\label{fig:xj_fraction}
\eef

We also estimated the component fractions of the dijet sample in p+p and 0\%-10\% Xe+Xe collisions at $\sqrt{s_{\rm NN}}=5.44$ TeV shown in \fig{fig:xj_fraction}, which would be helpful in understanding the role of jet flavor in jet-medium interactions. The left column shows the fractions of $gg$, $qg$, and $qq$ in the inclusive dijets in the p+p (top) and Xe+Xe (middle) collisions, as well as their differences (bottom). First, we find that $qg$ has the most prominent initial fraction ($\sim50\%$) in p+p collisions, and the fraction increases at a small $x_{\rm J}$ but decreases at $x_{\rm J}\sim 1$ in Xe+Xe collisions. Second, the fraction of $gg$ is overall reduced. At the same time, that of $qq$ is enhanced because, generally, the $gg$ dijets lose more energy, making it more difficult to survive the jet selection relative to $qq$. Because $qg$ is the most significant fraction in the dijet sample, the enhanced fractions of $qg$ at small $x_{\rm J}$ in Xe+Xe contribute to the increased $p_{\rm T}$ imbalance of the inclusive dijets. Furthermore, it is essential to address the fractional changes in the $q_{1}g_{2}$ and $g_{1}q_{2}$ subsets in the A+A collisions. In the right column of \fig{fig:xj_fraction}, it is observed that the fractions of $q_{1}g_{2}$ and $g_{1}q_{2}$ exhibit opposite behavior in the Xe+Xe collisions compared to their initial values. The fraction of $q_{1}g_{2}$ is significantly enhanced at small $x_{\rm J}$ after traversing the QGP, whereas that of $g_{1}q_{2}$ decreases in this region.

\bef
\includegraphics[width=0.9\linewidth]{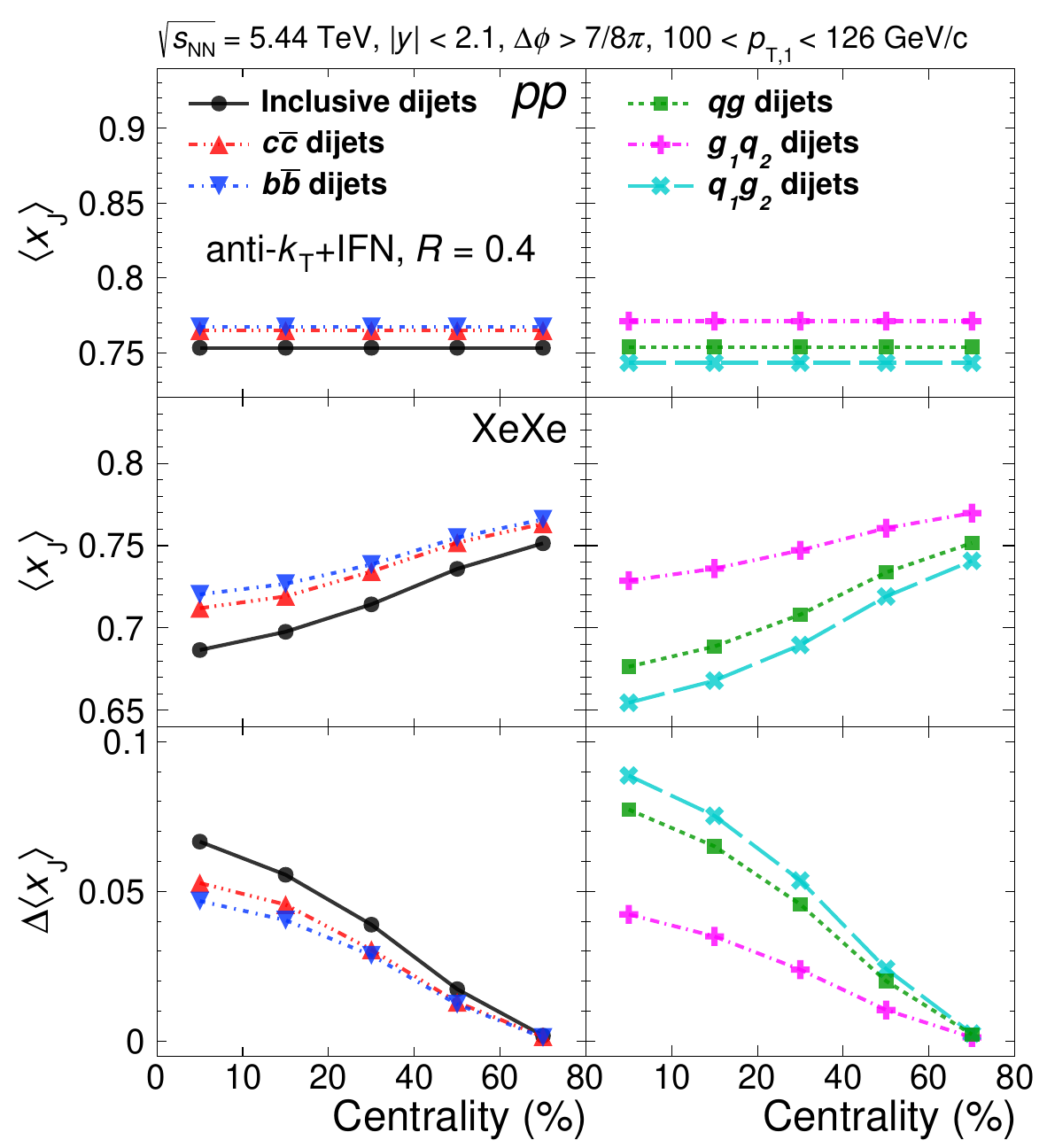}
\caption{ (Color online) Averaged $x_{\rm J}$ of the inclusive, $c\bar{c}$, $b\bar{b}$ dijets (left) and $q_{1}g_{2}$, $g_{1}q_{2}$, $qg$ dijets (right) in p+p (top) and Xe+Xe (middle) collisions as a function of the collision centrality, as well as their differences $\Delta \langle x_{\rm J} \rangle$ (bottom).}
\label{fig:delta_xj_cent_pp_XeXe}
\eef

To quantify the overall shift in the $x_{\rm J}$ distribution in Xe+Xe collisions relative to p+p, we calculated the average values ($\langle x_{\rm J} \rangle$) of dijet $x_{\rm J}$ distributions and their differences ($\Delta\langle x_{\rm J} \rangle$) between the p+p and A+A collisions, defined as follows:
\begin{align}
\langle x_{\rm J} \rangle &\equiv \int^1_0 \frac{1}{N_{\rm pair}} \frac{d N_{\rm pair}}{d x_{\rm J}} x_{\rm J} d x_{\rm J}, \\
\Delta \langle x_{\rm J} \rangle &= \langle x_{\rm J} \rangle_{\rm pp} - \langle x_{\rm J} \rangle_{\rm AA}.
\end{align}

To address the mass dependence of the medium modification of dijet $x_{\rm J}$ in the left column of \fig{fig:delta_xj_cent_pp_XeXe}, we show the $\langle x_{\rm J} \rangle$ of the inclusive, $c\bar{c}$ and $b\bar{b}$ dijets in p+p and Xe+Xe collisions as a function of centrality as well as their differences $\Delta \langle x_{\rm J} \rangle$. We find that $\Delta \langle x_{\rm J} \rangle$ decreases monotonously from central to peripheral collisions, as in \fig{fig:xj_5440_RAA}. It is also observed that the $\Delta \langle x_{\rm J} \rangle$ of these three kinds of dijets obey the hierarchy $\Delta \langle x_{\rm J} \rangle_{\rm incl.}>\Delta \langle x_{\rm J} \rangle_{\rm c\bar{c}}>\Delta \langle x_{\rm J} \rangle_{\rm b\bar{b}}$ in Xe+Xe collisions for the same centrality bin. This indicates that massive dijets suffer less asymmetric energy loss in A+A collisions than massless light flavors. Future measurements focusing on these comparisons will help test the mass effect of the jet energy loss. In the right column of \fig{fig:delta_xj_cent_pp_XeXe}, $\Delta \langle x_{\rm J} \rangle$ of $q_{1}g_{2}$ and $g_{1}q_{2}$ dijets in Xe+Xe are also plotted; the former has significantly larger values than the latter for each centrality bin.

\bef
\includegraphics[width=0.9\linewidth]{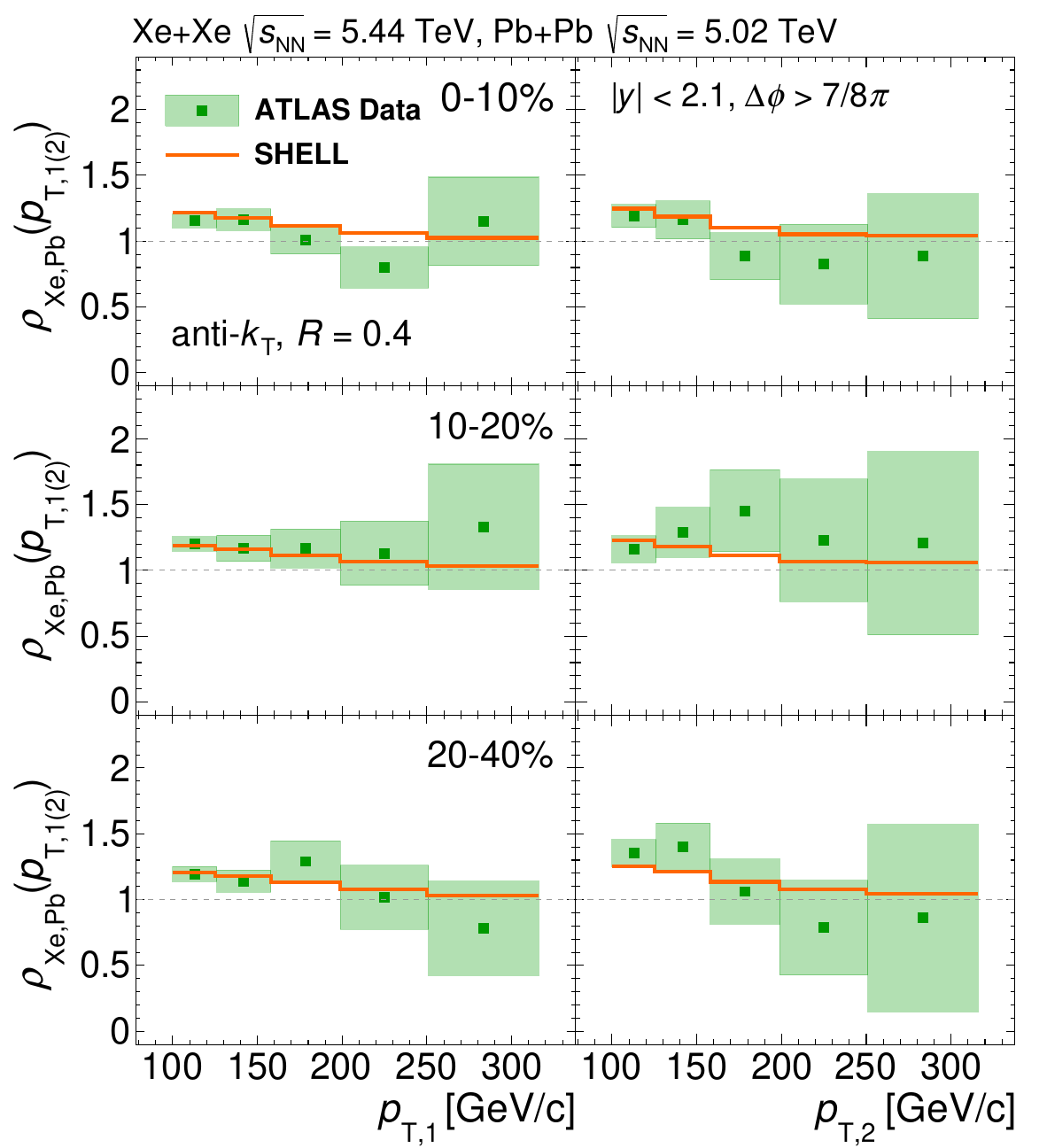}
\caption{ (Color online) Calculated ratios of Xe+Xe and Pb+Pb pair nuclear modification factors, $\rho_{\rm Xe,Pb}$, evaluated as a function of $p_{\rm{T,1}}$ (left) and $p_{\rm{T,2}}$ (right) in the same centrality intervals, and compared to the ATLAS data.}
\label{fig:rho_XePb}
\eef

To quantitatively characterize the relative dijet yield suppression between the Xe+Xe and Pb+Pb collisions, the nuclear modification factors for the leading jets, $\rho_{\rm Xe,Pb}(p_{\rm{T,1}})$, are defined as follows (similarly, $\rho_{\rm Xe,Pb}(p_{\rm{T,2}})$ can be defined for subleading jets):
\begin{align}
R_{\rm AA}(p_{\rm T,1}) &= \frac{1}{\langle T_{\rm AA} \rangle}
\frac{d\sigma^{\rm pair}_{\rm AA}/dp_{\rm T,1} }{
d\sigma^{\rm pair}_{pp}/dp_{\rm T,1}}, \\
\rho_{\rm Xe,Pb}(p_{\rm T,1}) &= \frac{R_{\rm XeXe}(p_{\rm T,1})}{R_{\rm PbPb}(p_{\rm T,1})}.
\label{eq:rho_XePb}
\end{align}

In \fig{fig:rho_XePb}, we show the calculated $\rho_{\rm Xe,Pb}(p_{\rm{T,1}})$ (left) and $\rho_{\rm Xe,Pb}(p_{\rm{T,2}})$ (right) of dijets for three centrality bins: 0\%-10\%, 10\%-20\% and 20\%-40\%. Note that we used a cut $x_{\rm J}>0.32$ in the calculation to be consistent with the ATLAS treatment in the measurements. We observe that the values of $\rho_{\rm Xe, Pb}$ are generally greater than one for both the leading and subleading jets for all centrality bins. This indicates a weaker yield suppression of dijets in Xe+Xe collisions than in Pb+Pb for the same centrality bin, and this phenomenon is still evident even in peripheral collisions. These findings were consistent with those of previous phenomenological studies \cite{Xie:2019oxg, Li:2021xbd, Zhang:2022fau}. Because the nucleus of xenon has a smaller radius than that of lead, the system size and mean temperature of the QGP medium formed in Xe+Xe collisions are expected to be smaller than those in Pb+Pb collisions within the same centrality interval \cite{Wang:2023udp}. Hence, dijets traverse a longer path length medium and experience more effective energy loss during Pb+Pb collisions.

\section{Conclusion}

In this paper, we present the first investigation of the medium modifications of dijet $p_{\rm T}$ balance ($x_{\rm J}$) in Xe+Xe collisions at $\sqrt{s_{\rm NN}}=5.44$ TeV. The initial $x_{\rm J}$ distributions of dijets were calculated using the POWHEG+PYTHIA8 prescription, which matched the NLO QCD matrix elements with the parton shower effect. The in-medium evolution of dijets in nucleus-nucleus collisions is described by the SHELL model, which considers both elastic and inelastic partonic interactions in the quark-gluon plasma (QGP). Our theoretical results for the dijet $x_{\rm J}$ in Xe+Xe collisions exhibit a more imbalanced distribution than in p+p collisions, consistent with the recently reported ATLAS data.
The dijet becomes increasingly imbalanced from peripheral to central Xe+Xe collisions, consistent with previous measurements of Pb+Pb collisions at the LHC. Furthermore, using an infrared-and-collinear-safe flavor jet algorithm, we explored the flavor dependence of the medium modification of dijet $p_{\rm T}$ balance in nucleus-nucleus collisions. We studied the respective medium-modification patterns and fraction changes of the $gg$, $qg$, and $qq$ components in the dijet sample for both p+p and Xe+Xe collisions. We demonstrate that the $qg$ component plays a key role in the increased imbalance in the dijet $x_{\rm J}$. In particular, we found that the $q_1g_2$ dijets experience a more significant asymmetric energy loss than the $g_1q_2$ dijets when traversing a QGP. By comparing the $\Delta \langle x_{\rm J}\rangle$ of inclusive, $c\bar{c}$ and $b\bar{b}$ dijets in Xe+Xe collisions, we observe $\Delta \langle x_{\rm J} \rangle_{\rm incl.}>\Delta \langle x_{\rm J} \rangle_{\rm c\bar{c}}>\Delta \langle x_{\rm J} \rangle_{\rm b\bar{b}}$ consistent with the mass hierarchy of partonic energy loss. In addition, the nuclear modification factors $\rho_{\rm Xe, Pb}$ of dijets in Xe+Xe at $\sqrt{s_{\rm NN}}=5.44$ TeV and Pb+Pb at $\sqrt{s_{\rm NN}}=5.02$ TeV are consistent with the ATLAS data, indicating that the yield suppression of dijets in Pb+Pb is more pronounced than that in Xe+Xe because of the larger radius of the lead nucleus.

\end{document}